# Ionic modulation at the LaAlO$_3$/KTaO$_3$ interface for extreme high-mobility two-dimensional electron gas

H. Yan[1$], S.W. Zeng[1$], K. Rubi[2,3*], G. J. Omar[1], Z. T. Zhang[1], M. Goiran[2], W. Escoffier[2], A. Ariando[1*]

[1]*Department of Physics, Faculty of Science, National University of Singapore, Singapore 117551, Singapore*

[2]*Laboratoire National des Champs Magnetiques Intenses (LNCMI-EMFL), CNRS-UGA-UPS-INSA, 143 Avenue de Rangueil, Toulouse 31400, France*

[3]*National High Magnetic Field Laboratory, Los Alamos National Laboratory, Los Alamos, NM 87545, USA*

[$]These authors contributed equally

*Email: rubi@lanl.gov; ariando@nus.edu.sg

**Due to the coexistence of many emergent phenomena, including 2D superconductivity and a large Rashba spin-orbit coupling, 5d transition metal oxides based two-dimensional electron systems (2DESs) have been prospected as one of the potential intrants for modern electronics. However, despite the lighter electron mass, the mobility of carriers, a key requisite for high-performance devices, in 5d-oxides devices remains far behind their 3d-oxides analogs. The carriers' mobility in these oxides is significantly hampered by the inevitable presence of defects generated during the growth process. Here, we report very high mobility (~ 22650 cm$^2$V$^{-1}$s$^{-1}$) of 5d-2DES confined at the LaAlO$_3$/KTaO$_3$ interface. The high mobility, which is beyond the values observed in LaAlO$_3$/SrTiO$_3$ and γ-Al$_2$O$_3$/SrTiO$_3$ systems in the same carrier-density range, is achieved using the ionic-liquid gating at room temperature. We postulate that the ionic-liquid gating affects the oxygen vacancies and efficiently reduces any disorder at the interface. Investigating density and mobility in a broad range of back-gate voltage, we reveal that the mobility follows the power-law $\mu \propto n^{1.2}$, indicating the very high quality of ionic-liquid-gated LaAlO$_3$/KTaO$_3$ devices, consistent with our postulate. Further, the analysis of the quantum oscillations measured in high magnetic fields confirms that the high-mobility electrons occupy the electronic sub-bands emerging from the Ta:5d orbitals of KTaO$_3$.**

**Introduction**

Because of its numerous intriguing characteristics, the two-dimensional electron gases (2DEGs) generated at the strongly correlated oxide interfaces have attracted significant interest in the field of condensed matter physics and are considered as a new candidate for electronic device applications[1,2]. The well-known and most widely explored oxide interface is the $LaAlO_3/SrTiO_3$ (LAO/STO) heterointerface that was discovered in 2004 to exhibit high mobility 2DEG[3], and subsequently shown to reveal fascinating properties such as magnetism, superconductivity and even coexistence of them[4-7]. Interestingly, the physical properties of the 2DEG at the LAO/STO interface are controllable through a back gate[8], a top gate[9], or an ionic liquid (IL) gate[10]. In particular, the electron mobility can be significantly enhanced by the IL gating, which triggers a switch of the 2DEG between a metallic and an insulating state[11]. To reveal the role of the Ti-ions to the emergent phenomena and obtain higher mobility 2DEG, many attempts to replace STO with other perovskite oxides have been made[12,13]. Among them, $KTaO_3$ (KTO), a polar and band-insulator ($E_g \sim 3.6$ eV) oxide, is particularly interesting[13-17]. Similar to STO, a 2DEG is realized at the metallic oxygen-deficient KTO surface[18] or the interface with other materials such as EuO/KTO[16], crystalline (*c*-) $LaVO_3$/KTO[15], amorphous (*a*-) LAO/KTO[15], and *c*-$LaTiO_3$/KTO[13]. As the origin of the electron gas is concerned, similar to STO-based interfaces, the origin of the 2DEG involves electrostatic relaxation and chemical reconstruction driven by the polar discontinuity[3], interface chemistry[18-20], and oxygen vacancies[21,22]. For a-LAO/KTO interface, the most likely scenario is the oxygen vacancies. In this scenario, the redox reaction at the interface, by oxidizing deposited LAO films and reducing the KTO substrate, accounts for the metallic interface. Furthermore, the gate-tunable 2D superconductivity is exclusively reported for the KTO (111) and (110) based interfaces[16,23]. Intriguingly, the large Rashba coefficient extracted from the spin-charge and charge-spin experiments makes KTO-2DEG a potential candidate for spin-orbitronic devices[23].

KTO is a 5*d* transition metal oxide that exhibits a lighter effective mass of electrons and a stronger spin-orbit coupling (SOC) at its conducting surface/interface than STO based interfaces. As a result of strong SOC, the low-lying Ta:$t_{2g}$ band splits at the Γ point and forms the light and heavy bands having the mixed character of $d_{xy}$, $d_{xz}$, and $d_{yz}$ orbitals[18,24,25]. Recent spectroscopy measurements, density functional theory calculations, and Shubnikov-de Haas oscillations[17] confirm the lighter effective mass of KTO-2DEG (~ 0.3-0.6 $m_e$)[18,26] compared to STO-2DEG (~ 0.6-1.8 $m_e$)[27]. Hence, KTO is envisioned as a superior platform for higher mobility 2DEG. Nevertheless, its experimental realization on as-grown interfaces has been hampered by disorders (such as oxygen vacancies, dislocations, or cation intermixing) present in the interfacial KTO layers. For example, the mobility of *c*-LaTiO$_3$/KTO grown by MBE is as low as 300 cm$^2$V$^{-1}$s$^{-1}$ at 2 K and EuO/KTO exhibits mobility of 111 cm$^2$V$^{-1}$s$^{-1}$ at 2 K[16].

In this letter, we demonstrate a significant enhancement of the mobility (from 6000 to 22650 cm$^2$V$^{-1}$s$^{-1}$ at $T = 3$ K) of 2DEG at the *a*-LAO/KTO interface using IL gating at room temperature. In oxide-based heterostructures, electrons scattering by oxygen vacancies hinders their mobility. The IL gating partially fills the oxygen vacancies and enhances the electrons' mobility. The effect of different partial oxygen pressure during the growth of *a*-LAO is also studied and is found to have only minimal influence. We also investigated the back-gate influence on the carrier density, mobility, and quantum oscillations at low temperatures by measuring high field magneto-transport on our best-quality *a*-LAO/KTO device. The large-amplitude quantum oscillations and a linear dependence of mobility with respect to the carrier density further confirm the high quality of the 2DEG in IL gated *a*-LAO/KTO devices.

**Results and discussion**

**Ionic-liquid gating**

The *a*-LAO/KTO devices were fabricated by depositing a 4 nm *a*-LAO thin film on KTO (100) substrate at different oxygen partial pressures ($P_{O_2}$ =10$^{-3}$, 10$^{-4}$, 10$^{-5}$, and 10$^{-6}$ Torr) in the pulsed laser deposition chamber. Before a-LAO deposition, a Hall-bar geometry was structured at the KTO surface using an AlN masking layer and conventional photolithography technique. An isolated gate electrode was created at the side of the Hall bar for IL gating. Figure 1(a) depicts a schematic of the a-LAO/KTO microstructure, where a droplet of IL covers both the conduction channel and the gate electrode. Figures 1(b) and 1(c) show the resistance of the device ($R = V_{SD}/I_{SD}$) as a function of the IL gate voltage ($V_{ILG}$) over four cycles for two devices, where the a-LAO thin film was deposited at oxygen partial pressure of $1\times10^{-6}$ and $1\times10^{-5}$ Torr, respectively.

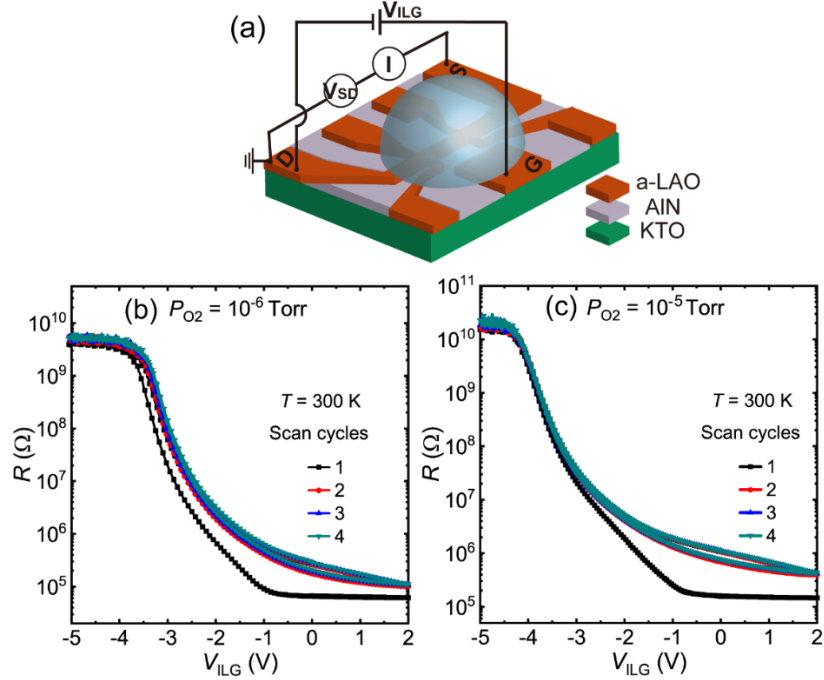

**Fig. 1 (One-column figure): An ionic-liquid (IL) gate sweep on *a*-LAO/KTO interface at room temperature.** (a) A schematic of the *a*-LAO/KTO microstructure where the IL covers the *a*-LAO channel and the gate electrode. S is a source contact, D is a drain contact, and G denotes a gate electrode. (b), (c) Typical resistance $R$ as a function of IL gate voltage ($V_{ILG}$) for *a*-LAO/KTO for which the *a*-LAO film with a thickness of 4 nm is deposited at $P_{O_2}$ = $1\times10^{-6}$ Torr and $1\times10^{-5}$ Torr, respectively.

Each $V_{ILG}$ scan cycle starts from 0 to 2 V, and then from 2 to -5 V, and finally, from -5 to 0 V with a scan speed of 25 mV/s. In the first scan cycle (0 → +2 → -5 → 0 V), both devices switch from low (~ $10^5$ Ω) to high-resistance (~ $10^{10}$ Ω) states irreversibly. However, the sample's resistance is reversible with weak hysteresis for subsequent cycles. This effect arises from the irreversible filling of oxygen vacancies during the first cycle of electrolyte gating due to oxygen electro-migration from the *a*-LAO layer[28,29]. After completing a few cycles, we set the gate voltage to $V_{ILG} = 0$ V, clean off the IL and then start cooling down the devices for electrical transport measurements at low temperature. Similar transport measurements on the pristine devices, *i. e.* before IL gating, are also carried out and used as a reference.

We compare the temperature dependence of sheet resistance ($R_s$) of the *a*-LAO/KTO device ($P_{O_2} = 1\times10^{-6}$ Torr) before and after IL gating in Fig. 2(a). The metallic nature is preserved and improved after gating with higher and lower $R_s$ at high and low temperatures (< 10 K), respectively. It is worth mentioning here that the *a*-LAO/KTO device ($P_{O_2} = 1\times10^{-2}$ Torr) was highly insulating even before the IL gating process, consistent with previous reports[22]. We, however, observed similar results before and the after IL gating effect on *a*-LAO/KTO device, for which *a*-LAO is grown at three other different $P_{O_2}$ ($10^{-3}$, $10^{-4}$, $10^{-5}$ Torr). The lower $R_s$ at low temperature after IL gating is most likely due to the decreasing of the density of oxygen vacancies which act as scattering centers and cause the relatively high resistance before gating[22,30]. A similar effect has also been reported for the 2DEG at the *a*-LAO/STO interface[29]. The striking influence of IL gating is evident in the Hall resistance variations, as shown in Figs. 2(b) and 2(c). In Fig. 2b, we show the magnetic field (*B*) dependence of $R_{xy}$ before IL gating measured at a few selected temperatures ranging from 100 to 2.5 K. When the temperature is above 100 K, $R_{xy}$ varies linearly with the magnetic field (not shown here), while it turns non-linear when the sample is cooled at temperatures below 100 K. Differently, $R_{xy}$ varies linearly with *B* in the whole temperature range from 300 to 2.5 K after IL gating (Fig. 2c). To extract the Hall carrier density $n_H$ and the Hall mobility $\mu_H$, we fit the non-linear $R_{xy}(B)$ data with the two-band model (solid lines in Fig. 2b)[31], whereas the linear traces are treated with a single carrier model (dashed lines). The room-temperature density of charge carriers after gating is around $\sim 3 \times 10^{13}$ cm$^{-2}$, much lower than that ($\sim 1.6 \times 10^{14}$ cm$^{-2}$) before gating, suggesting that part of oxygen vacancies is filled during the IL gating process. After gating, $n_H$ is nearly temperature independent with a slight upturn at low temperature, which is in contrast to the behavior of the device before gating, for which $n_H$ decreases by one order of magnitude from 300 to 2.5 K. The decrease in $n_H$ with lowering temperature implies the carrier freeze-out effect characterized by activation energy $E_a$ (the energy difference between oxygen-vacancies donor level and the minimum of conduction band). By fitting the temperature-dependent carrier density before-gating with the Arrhenius law $n \propto e^{-E_a/k_B T}$, we estimate $E_a \sim 12$ meV. Further, the absence of carriers' freeze-out after gating reveals the filling of a part of oxygen vacancies and a significant reduction of the gap between donor states and the minimum of the conduction band, consistent with the IL gating effect on $a$-LAO/STO[29,32]. The filling of the oxygen vacancies induces a significant enhancement of the mobility after gating. Impressively, the mobility increases from 5590 to 11800 cm$^2$/Vs at 2.5 K after IL gating, inferring the enhanced ratio of 111%.

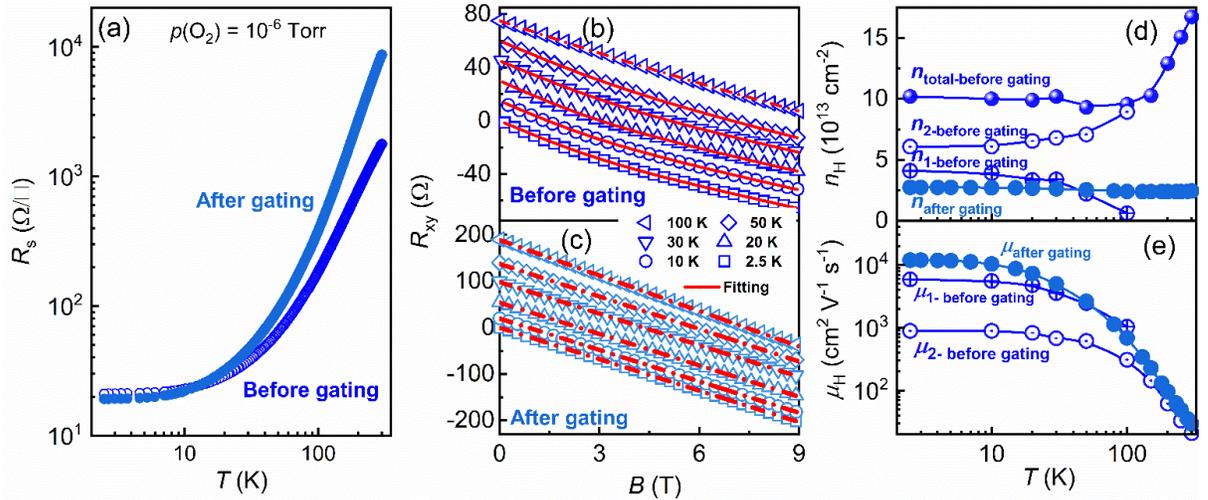

**Fig. 2 (Two-column figure): A comparison of electrical properties of $a$-LAO/KTO ($10^{-6}$ Torr) before and after the ionic-liquid gating process.** (a) Temperature dependence of the sheet resistance. (b) and (c) Magnetic field dependence of Hall resistance at various temperatures down to 2.5 K. Symbols are the experimental data, solid lines are the fit to the two-band model, and the dash lines are the fit to the single-band model. The experimental as well as the fitted data are shifted along the y-axis for clarity. (d) Carrier density and (e) mobility extracted from the Hall resistance data fitting at different temperatures. After the ionic-liquid gating process, the carrier density remains almost unchanged with temperature.

To investigate the growth oxygen partial pressure ($P_{O_2}$) effect on the mobility enhancement by IL gating, we compare four $a$-LAO/KTO devices for which the $a$-LAO was deposited under different values of $P_{O_2}$ ($10^{-3}$, $10^{-4}$, $10^{-5}$, and $10^{-6}$ Torr), after IL gating. Figures 3(a), 3(b), and 3(c) display $R_s$, $n_H$, and $\mu_H$, respectively, as a function of temperature. For each device, a metallic behavior is observed after gating. Among them, the sample grown at $10^{-4}$ Torr surprisingly exhibits the best metallicity with the lowest $R_s$ about 8 Ω/□ at 3 K. The $n_H$ of the samples after gating show nearly $T$ independent behavior in the range of $2.55 \times 10^{13}$ to $3.75 \times 10^{13}$ cm$^{-2}$ and do not exhibit a significant and progressive $P_{O_2}$ dependence. The room-temperature $\mu_H$ is 26-33 cm$^2$V$^{-1}$s$^{-1}$, which is higher than that in STO-based interfaces (which is usually below 10 cm$^2$V$^{-1}$s$^{-1}$)[3] and similar to previous KTO-based interfaces, e.g., LaTiO$_3$/KTO (21 cm$^2$V$^{-1}$s$^{-1}$)[13]. The $\mu_H$ increases with the decreasing temperature down to 2.5 K after IL gating, and the highest value for $\mu_H$ (22650 cm$^2$V$^{-1}$s$^{-1}$) was obtained for the sample grown at $P_{O_2} = 10^{-4}$ Torr. Overall, the high mobility obtained by filling the oxygen vacancies (O$_v$) as a result of IL gating is not significantly affected by $P_{O_2}$. The mobility enhancement in the present results is much higher than that in IL/STO (~1000 cm$^2$V$^{-1}$s$^{-1}$) and IL/KTO (~7000 cm$^2$V$^{-1}$s$^{-1}$) interfaces[33,34].

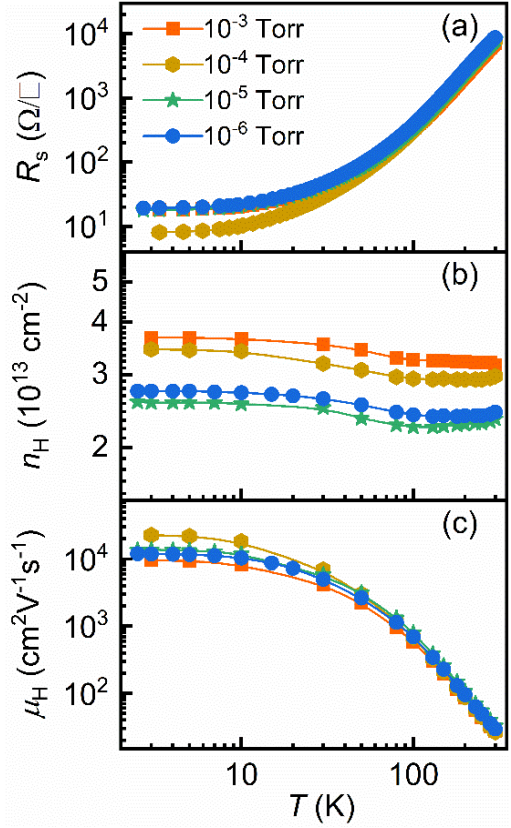

**Figure 3 (one-column figure): Growth oxygen partial pressure ($P_{O_2}$) effect on the electrical properties of *a*-LAO/KTO after the ionic-liquid gating.** Temperature dependence of (a) sheet resistance $R_s$, (b) Hall density $n_H$, and (c) Hall mobility $\mu_H$. The sample for which the *a*-LAO was deposited at $P_{O_2} \sim 10^{-4}$ Torr shows the highest mobility ~ 22650 cm$^2$V$^{-1}$s$^{-1}$ at low temperatures.

**Back gating and quantum oscillations**

To further examine the quality of IL gated *a*-LAO/KTO devices, we investigate the back-gate effect on carrier density and mobility at low temperatures. For this purpose, we measure the longitudinal and Hall resistance of one device ($P_{O_2} = 10^{-4}$ Torr) in a high pulsed field (53 T) and at various back-gate voltages ($V_{BG}$) ranging from +100 V to -125 V and applied across the KTO substrate. It is worth mentioning that these measurements were carried out several months after the IL gating process. The scheme of transport measurements with the back gate is displayed in the inset of Fig. 4(a), while $R_{xx}(B)$ and $R_{xy}(B)$, for which the magnetic field was parallel to the normal of KTO (100) plane, are shown in the main panels of Fig. 4(a) and Fig. 4(b), respectively.

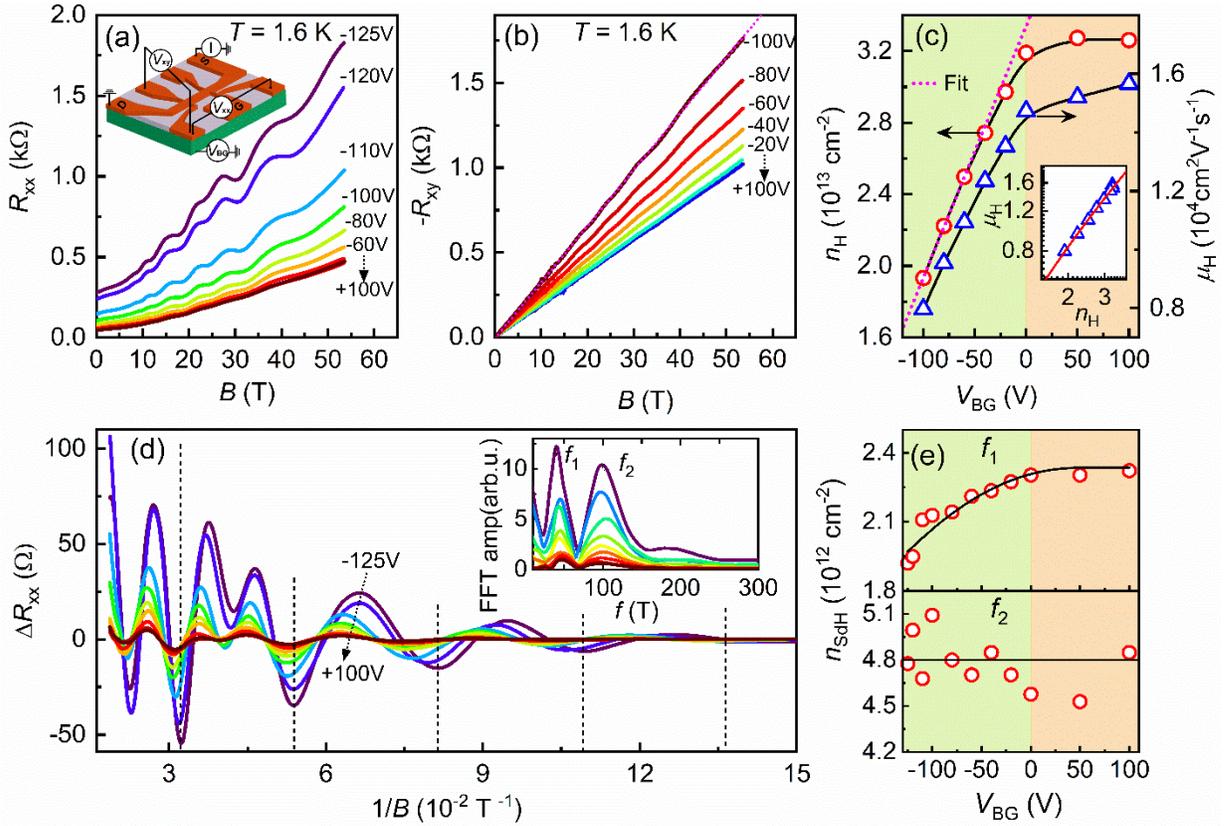

**Fig. 4 (Two-column figure): Back-gate evolution of electrical properties and quantum oscillations in the $a$-LAO/KTO ($P_{O_2}$ ~ $10^{-4}$ Torr) sample.** (a) Longitudinal resistance and (b) Hall resistance as a function of the field for different back-gate voltage $V_{BG}$ ranging from -125 V to +100 V. Inset of (a) shows the scheme of transport measurements with back gate. The magnetic-field direction was parallel to the normal of KTO (100) plane. The dashed line in (b) is the linear fit to the Hall resistance. (c) Main panel: Variation in Hall carrier density $n_H$ (left y-axis) and Hall mobility $\mu_H$ (right y-axis) as a function of $V_{BG}$. A parallel-plate capacitor model fit to the $n_H$ data is displayed with a dashed line. Inset: $\mu_H$ as a function of the $n_H$ with linear fit displayed as a solid line. The units of $n_H$ and $\mu_H$ are the same as in the main panel. (d) Inverse-field dependence of the oscillating longitudinal resistance $\Delta R_{xx}$. Inset shows the fast Fourier transform (FFT) analysis. (e) The carrier density calculated from the oscillations' frequencies $f_1$ (upper panel) and $f_2$ (lower panel). Solid lines in the main panel of (c) and (e) are guides to the eyes.

We note that the Hall voltage was not measurable at $V_{BG}$ < -100 V since the 2DEG turns to be inhomogeneous at such a high negative $V_{BG}$, similar to the 2DEG at the LAO/STO interface. We, however, do not observe any irreversibility or hysteresis while sweeping the $V_{BG}$ back and forth, as reported for the STO-2DEG[35]. Large-amplitude quantum oscillations, superimposed on a positive magnetoresistance background, are perceived for $B$ > 6 T (Fig. 4(a)). On the other hand, $R_{xy}$ (B) remains linear in the full-field range with a few weak quantum oscillations. With increasing $V_{BG}$ from 0 V to +100 V, no significant difference in $R_{xx}(B)$ and $R_{xy}(B)$ is observed, however, the positive magnetoresistance and the oscillations' amplitude meaningly enhance with varying $V_{BG}$ from 0 V to -125 V. The enhancement of $R_{xx}(B = 0\text{ T})$ certainly indicates the depopulation of charge carriers at the interface as expected for negative values of $V_{BG}$. This effect becomes even clearer when we plot the Hall carrier density $n_H$ as a function of $V_{BG}$ on the left y-axis of Fig. 4(c), where the saturation of $n_H$ in the positive $V_{BG}$ regime is similar to that observed for the 2DEG at the LAO/STO interface[35,36]. It is probable that beyond a certain $V_{BG}$ value, any additional charge carriers start escaping the quantum well formed at the interface and are trapped by defects into the bulk KTO. For $V_{BG} < 0$ V, $n_H$ first drops marginally but follows a rapid and linear decrease below $V_{BG} = -40$ V. From the parallel-plate capacitance model $\delta n = \frac{\varepsilon_0 \varepsilon_r}{et} \delta V_{BG}$, where $\delta n$ is the change in density, $\varepsilon_0$ is the vacuum permittivity, $\varepsilon_r$ (= 5000[36]) and $t$ (= 500 µm) are the dielectric permittivity and thickness of KTO, respectively, only 17% change in charge carrier density is expected for $\delta V_{BG} = 100$ V, while we experimentally observe a change of 40%. This discrepancy may find an explanation in the filling of some oxygen vacancies at the interface by applying the negative $V_{BG}$. Contrary to our results, a smaller back-gating effect observed on a higher density a-LAO/KTO device[37] was attributed to electric-field modulation of the effective disorders at the interface[38].

For negative back-gate voltage, the Hall mobility $\mu$ decreases steeply from 14720 cm$^2$V$^{-1}$s$^{-1}$ (0 V) to 8000 cm$^2$V$^{-1}$s$^{-1}$ (-100 V) as shown in the right y-axis of Fig. 4(c). Contrary to the back-gate evolution of the mobility in as-grown a-LAO/KTO(111)[37], the mobility follows a power law $\mu \propto n^\alpha$ with exponent $\alpha = 1.2$ (see inset Fig. 4(c)), which is comparable to that for very high-mobility 2DEGs in γ-Al$_2$O$_3$/STO[39] and semiconductor heterostructures[40,41]. This large exponent indicates that the electron scattering in the IL gated a-LAO/KTO devices is most likely dominated by the remote ionized impurity (e. g. oxygen vacancies), rather than disorder inside the quantum well. We also notice that the mobility of charge carriers degrades with time after the IL gating process, as the samples are stored in ambient conditions. Since the measurements in the high magnetic field are carried out several months after the IL gating process, the mobility is lower than the previously recorded maximum value of 22650 cm$^2$V$^{-1}$s$^{-1}$, but remains higher than that of the pristine interface.

Next, we focus on the back-gate evolution of quantum oscillations observed in the longitudinal resistance after subtracting a smooth background. The oscillating part of the magneto-resistance $\Delta R_{xx}$ is displayed in Fig. 4(d) as a function of the inverse magnetic field. When $V_{BG}$ increases, the higher Fermi energy with the band-structure affects the set of frequencies $f_i$, which composes the quantum oscillations. This effect is perceived as a shift of the oscillations' extremum towards lower values of the inverse magnetic field. Contrary to previous work where four frequencies are resolved[17], the fast Fourier transform analysis of $\Delta R_{xx}(1/B)$ yields only two frequencies for all $V_{BG}$ (see inset of Fig. 4(d)). For $V_{BG} = 0$ V, $f_1 = 47$ T and $f_2 = 100$ T. Since $f_2 \sim 2f_1$, it is reasonable to assume that the spin degeneracy is resolved at high fields ($B > 20$ T) owing to the Zeeman effect. For an in-depth investigation of this assumption, we simultaneously analyze the inverse-field dependence of the quantum oscillations in $R_{xx}$ and $R_{xy}$. We intentionally used the data for $V_{BG} = -100$ V because the oscillations' amplitude for both $R_{xx}(B)$ and $R_{xy}(B)$ is larger in the low-density regime. Not to entirely rely on the results from the FFT analysis, we simulate quantum oscillations in $R_{xx}(B)$ and $R_{xy}(B)$, considering a population of two subbands and using the Lifshitz-Kosevich (L-K) equation as given below[45]

$$\Delta R_{xx} = \sum_{i=1,2} R_{0i} R_{Ti} R_{Di} \cos\left[\frac{2\pi F_i}{B} + \frac{3\pi}{4}\right] \quad (1)$$

$$\Delta R_{xy} = \sum_{i=1,2} R_{0i} R_{Ti} R_{Di} \cos\left[\frac{2\pi F_i}{B} + \frac{\pi}{4}\right] \quad (2)$$

where $R_0$ is magnetic-field-independent resistivity. $R_T$ and $R_D$ are the temperature- and field-damping factors expressed as $R_T = \frac{2\pi^2 k_B T m_c/\hbar eB}{Sinh(2\pi^2 k_B T m_c/\hbar eB)}$, $R_D = exp\left(-\frac{2\pi^2 k_B T_D m_c}{\hbar eB}\right)$. $T_D$ is the Dingle temperature related to the quantum relaxation time $\tau$ as $T_D = \frac{\hbar}{2\pi k_B \tau}$.

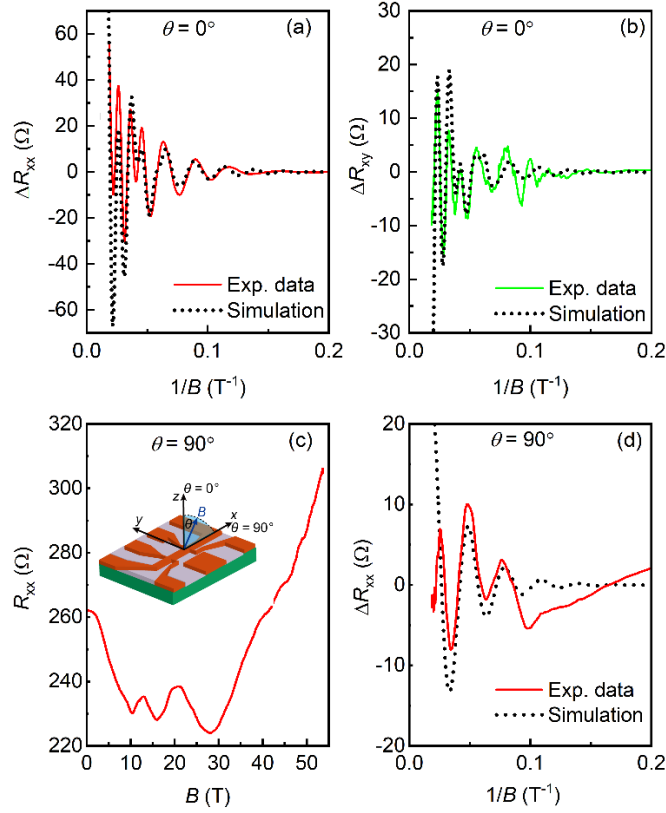

**Fig. 5. (One-column figure): Comparison of experimental and simulated oscillations for different magnetic-field orientations.** Inverse-field dependence of oscillatory resistances (a) $\Delta R_{xx}$ and (b) $\Delta R_{xy}$ at $\theta = 0°$, $V_{BG} = -100$ V, and $T = 1.6$ K. (c) $R_{xx}(B)$ and (d) $\Delta R_{xx}(1/B)$ for $\theta = 90°$ measured at $V_{BG} = -100$ V and $T = 1.6$ K. Inset in (c) displays magnetic field orientation with respect to the KTO (100) plane for $\theta$ values. Solid and dotted lines are the experimental and simulated data, respectively. $\Delta R_{xy}$ is estimated by subtracting the linear-fit data from the $R_{xy}(B)$. The non-monotonous magnetic-field dependence of $R_{xx}$ in (c) leads to an imperfect background subtraction for $B < 12$ T.

We compare the simulated and experimental $\Delta R_{xx}$ and $\Delta R_{xy}$ data in Fig. 5(a) and (b), respectively. As expected from the relation $R_{xx} \propto \frac{dR_{xy}}{dB} \times B$ [43], the oscillations in $R_{xy}$ show a phase shift of $\pi/2$ with oscillations in $R_{xx}$. For $B > 10$ T, the simulated data agree well with the experimental data for $f_1 = 41$ and $f_2 = 100$ T, except for the deviation in oscillations' amplitude that can be corrected by choosing appropriate $R_{0i}$ and $T_{Di}$ values. The discrepancy below 10 T is likely because of the poor resolution of oscillations in small fields. It is worth mentioning that from the FFT analysis of $\Delta R_{xx}(1/B)$ for $V_{BG}$ = -100 V, we obtained the frequencies of 43 T and 100 T, resulting in a 5% error in $f_1$. Since the oscillations in both $R_{xx}(B)$ and $R_{xy}(B)$ are resonant in terms of frequencies, we can simply discard any significant error induced from the background subtraction that could lead to the shift of a few Teslas in $f_1$ or $f_2$ and, therefore, the presence of oscillations from the spin-split subbands in high magnetic fields. The charge carrier density $n_{SdH}$ $\left(=\frac{2e}{h}f\right)$ calculated from both frequencies are plotted as a function of $V_{BG}$ in the upper and lower panels of Fig. 4(e). While $n_{SdH}(f_1)$ mimics the behavior of $n_H$, we do not observe any clear trend for the variation of $n_{SdH}(f_1)$ with $V_{BG}$. Interestingly, the $n_{SdH}(f_1)$ changes only ~ 17 % with varying $V_{BG}$ from 0 to -100 V as expected from the parallel-plate capacitance model when $\varepsilon_r$ (KTO) is 5000. We estimate the total carrier density from the estimated frequencies as $n_{SdH} = \frac{2e}{h}\sum f_i =$ $7.1 \times 10^{12}$ cm$^{-2}$, which is much smaller than the $n_H = 3.2 \times 10^{13}$ cm$^{-2}$ at $V_{BG} = 0$ V. This difference can be persuasively explained by the fact that $n_{SdH}$ does not include the contribution of populated sub-bands having low charge carrier mobility at 1.6 K. This discrepancy, however, may also find an explanation in the presence of numerous parallel 2D conducting subbands and/or the coexistence of 2D and 3D carriers. In the present case, the first scenario is most unlikely because the subbands for KTO-DEG predicted by theoretical calculations and supported by the results from the angle-resolved photoemission spectroscopy[18,44] and SdH oscillations measured at very low temperatures[17] do not agree with at least four parallel conducting subbands of the same density and effective masses, like the ones predicted in $a$-LAO/La$_{7/8}$Sr$_{1/8}$MnO$_3$/STO heterostructures[42]. To investigate the existence of 3D carriers and their implication for the discrepancy in $n_H$ and $n_{SdH}$, we measured the same sample at $V_{BG}$ = -100 V and $\theta = 90°$ (magnetic field orientated parallel to the KTO (100) as displayed in the inset of Fig. 5(c)). We show the best results of the subtracted oscillatory $R_{xx}$ as a function of the inverse magnetic fields in Fig. 5(d). To avoid using distrustful results from the FFT of two oscillations, we simulate the oscillations using Eq. (1) and show them as a dotted line in Fig 5(d) along with the experimental data (solid line). The simulated data reasonably agree with the experimental data when the oscillations frequency is 33 T, which translates into the 3D-SdH density of $1.08 \times 10^{18}$ cm$^{-3}$ as $n_{SdH}^{3D} = \frac{8}{3\sqrt{\pi}} \left(\frac{ef}{h}\right)^{3/2}$. By comparing the $n_{SdH}$ for 3D carriers and the residual 2D carriers' density ($n_{Hall}^{2D} - n_{SdH}^{2D} = 2.5 \times 10^{13}$ cm$^{-2}$), we estimate the depth of the electron gas $\left(t = \frac{n_{2D}}{n_{3D}}\right)$ of ~ 23 μm, which seems implausible after comparing the electron gas depth of ~ 20 nm for a similar value of Hall density of ionic-liquid gated KTO surface[34]. We, therefore, conclude that the 3D carriers share the residual 2D carriers' density only partially and do not fully explain the discrepancy between $n_{SdH}$ and $n_H$.

To estimate the cyclotron mass $m_c$, the temperature-dependent FFT amplitude of the quantum oscillations at $V_{BG} = 0$ V (Fig. 6) was fitted to the standard Lifshitz-Kosevich (L-K) formula[45] given below.

$$A(T) = A_0 \frac{2\pi^2 k_B m_c T / \hbar e B_{eff}}{\sinh(2\pi^2 k_B m_c T / \hbar e B_{eff})} \quad (3)$$

where $\frac{1}{B_{eff}} = \frac{\frac{1}{B_{min}} + \frac{1}{B_{max}}}{2}$, $A_0$ is constant, $\hbar$ is the reduced Planck constant, $k_B$ is the Boltzmann constant, $B_{min}$ and $B_{max}$ are the magnetic field bounds, in between of which the FFT is calculated. The best-fitting as shown in Fig. 6(c) yields $m_c$ values of $0.52 \pm 0.02$ $m_e$ ($f_1 = 47$ T) and $0.67 \pm 0.03$ $m_e$ ($f_2 = 100$ T), which corresponds to the heavy electron mass in the 2D-confined electron system of KTO[17,44]. Similar to LAO/STO, we conclude that the contribution to the quantum oscillations expected from the sub-bands occupying the lighter electrons are not resolved in *a*-LAO/KTO (for the experimentally investigated temperature range), even though the estimated cyclotron mass is three times smaller[36,46,47]. The unresolved oscillations find an origin in low mobility of carriers populating the light subbands. Since the light sub-bands has a strong $d_{xy}$ orbital character mostly extending in the interface-adjacent $TaO_2$ planes[44,48], the mobility of carriers populating these bands is most likely impeded by the substantial disorders (e. g. dislocations[14,16], mixed valance states of Ta ions[49], and substitution of La ions at the K sites[16]) in the interface-adjacent KTO layers.

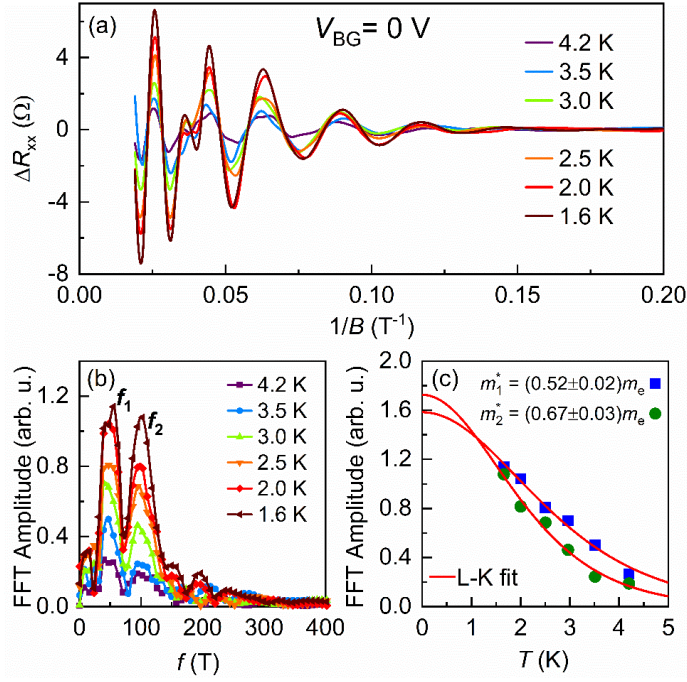

**Fig. 6 (One-column figure): Analysis of quantum oscillations measured at different temperatures and at $V_{BG} = 0$ V for the *a*-LAO/KTO ($P_{O2} \sim 10^{-4}$ Torr) sample.** (a) Inverse-field dependence of the oscillating resistance $\Delta R_{xx}$ obtained after subtracting the non-oscillating background from the measured longitudinal resistance. (b) Fast Fourier transform (FFT) amplitude as a function of frequency. Two peaks at frequencies $f_1$ = 47 T and $f_2$ = 100 T are observed at all temperatures. (d) Temperature dependence of the FFT amplitude for both frequencies. Symbols are the experimental data, and the solid lines are the Lifshitz-Kosevich (L-K) fit.

**Conclusion**

In summary, in an effort to improve the quality of oxides-based 2DEG, we investigate IL gating effects on the transport properties of the *a*-LAO/KTO interface. After IL gating at room temperature, we observe a ~ 200 % enhancement in the mobility of charge carriers. The highest mobility (~ 22650 $cm^2V^{-1}s^{-1}$) measured at low temperature is even higher than the record value ($\sim$11000 cm$^2$V$^{-1}$s$^{-1}$) reported in the oxide interface γ-Al$_2$O$_3$/STO[50], in the same carrier density regime ($\sim$ 10$^{13}$ cm$^{-2}$). The large amplitude of Shubnikov-de Haas oscillations and the linear carrier-concentration dependence of the mobility authenticate the high quality of IL-gated *a*-LAO/KTO devices. We conclude that the mobility of 2DEG at the KTO based interfaces is mainly limited by the inevitable presence of oxygen vacancies. Our results demonstrate that the control of oxygen vacancies through the IL gating process and the lighter effective mass of electrons in *a*-LAO/KTO systems lead to higher mobility of 2DEG at the interface compared to the LAO/STO interface. The significant enhancement of the 2DEG mobility at the *a*-LAO/KTO interfaces opens new possibilities to progress further in the realization of new-generation oxide-based electronic and quantum devices.

**Methods**

**Growth and device fabrication.** The *a*-LAO/KTO heterostructures were obtained by depositing 4 nm *a*-LAO thin films on (100)-oriented KTO substrates using a pulsed laser deposition (PLD) system. The LAO target was a single crystal. The deposition temperature was 25°C and oxygen partial pressure $P_{O_2}$ was from 10$^{-3}$ to 10$^{-6}$ Torr. Before deposition, patterns with a six-terminal Hall bar device and a lateral gate electrode were fabricated on KTO by depositing *a*-AlN films as a hard mask and using conventional photolithography. In order to remove the oxygen vacancies introduced in KTO during deposition of *a*-AlN, the patterned KTO were annealed in a tube furnace at 500 °C in air for 1 hour. The width of the Hall bar is 50 μm and the distance between two voltage probes is 160 μm.

**Gating and magnetotransport measurements.** The electrical connections for transport measurement were made using ultrasonic wire bonding. A small drop of the IL, *N*,*N*-diethyl-*N*-methyl-*N*-(2-methoxyethyl)ammonium bis(trifluoromethyl sulphonyl)imide (DEME-TFSI), covered both the Hall-bar channel and the gate electrode. The gating experiments were performed at room temperature using the Keithley 2400 Sourcemeters and 2002 Multimeters. The electrical transport measurements in a moderate field regime (0-9 T) were performed in Physical Property Measurement System (PPMS) from Quantum Design. The transport measurements in high magnetic fields were performed in a $^4$He cryostat at the Laboratoire National des Champs Magnetiques Intenses (LNCMI) in Toulouse, France. The longitudinal and Hall voltages were simultaneously measured in a high pulsed magnetic field up to 53 T with a duration of 300 ms. The excitation current for high-field magnetotransport measurements was 1 µA.

**Acknowledgments**

This research is supported by the Agency for Science, Technology, and Research (A*STAR) under its Advanced Manufacturing and Engineering (AME) Individual Research Grant (IRG) (A1983c0034). We acknowledge the support of LNCMI-CNRS, member of the European Magnetic Field Laboratory (EMFL). A. Ariando acknowledges the National Research Foundation (NRF) of Singapore under its NRF-ISF joint program (Grant No. NRF2020-NRF-ISF004-3518) for the financial support. K. Rubi acknowledges support from the National High Magnetic Field Laboratory, supported by the National Science Foundation through NSF/DMR-1644779 and the State of Florida, and the US Department of Energy "Science of 100 Tesla" BES program.